\documentstyle[eqsecnum,aps]{revtex}
\begin{document}
\title{Transitions between complex configurations in the excitation of the Double
Giant Resonance}
\author{V.Yu. Ponomarev$^{1,2}$\thanks{%
e-mail: vlad@thsun1.jinr.ru} and C.A. Bertulani$^{1}$\thanks{%
e-mail: bertu@if.ufrj.br}}
\address{$^1$Instituto de F\'\i sica, Universidade Federal do Rio de Janeiro,
21945-970 Rio de Janeiro, RJ, Brazil }
\address{$^2$Bogoliubov Laboratory of Theoretical
Physics, 
Joint Institute for Nuclear Research, 141980, Dubna, Russia}
\date{\today }
\maketitle

\begin{abstract}
The transitions between complex configurations, to which the giant dipole
resonance (GDR) and the double giant dipole resonance (DGDR) doorway states
are coupled, are taken into account in second order perturbation theory for
the reaction amplitude. It is proved that only transitions between GDR and
DGDR doorway states play an essential role in the Coulomb excitation of the
DGDR.
\end{abstract}

%\draft
%\preprint{Draft}

\bigskip

\noindent
PACS numbers: 23.20.-g, 24.30.Cz, 25.70.De, 25.75.+r

\bigskip

%\narrowtext
\twocolumn

After the first observation of the double giant dipole resonance (DGDR) in
relativistic heavy ion collisions (RHIC) \cite{Rit93,Sch93} the magnitude of
its excitation cross section attracts a special attention. This is because
the first data \cite{Sch93} indicated a strong enhancement of the DGDR
excitation in $^{136}$Xe (by a factor of 2-4) as compared to the theoretical
predictions available \cite{Ber93,Pon94}. Several attempts have been made to
understand the reason of this phenomena \cite{Ber96a,Ber96,Lan97,Bor97,Pon97}%
. A few years later a similar experiment with the excitation of the DGDR in $%
^{208}$Pb brought another intriguing news \cite{Bor96} which were
interpreted as a good agreement between experiment and theory if some
corrections to the pure harmonic picture of the DGDR excitation were added 
\cite{Ber96,Lan97,Bor97,Pon97}. Unless a good systematic study is achieved,
the question on the excitation cross section of the DGDR in RHIC remains
open. This stimulates theoretical studies on the different processes which
might be responsible for an enhancement of the DGDR strength function.

Recently it was argued that transitions between complex configurations, to
which the GDR and DGDR doorway states are coupled, may be responsible for
the enhanced DGDR decay into the GDR states as compared to the GDR decay
into the ground state \cite{Sol97}. However, it should be remembered that
because of the available phase space two consequent $\gamma $-emissions from
the DGDR is not the same as the inverse process of the DGDR excitation. The
transitions between the GDR and the DGDR complex configurations were not
taken into account in previous microscopic studies \cite{Pon97,Pon96a,Pon96b}%
. The role of these transitions for the DGDR excitation in RHIC will be
considered in the present paper. It will be concluded that their role is
marginal in the process under consideration although a huge amount of the $%
E1 $-strength is hidden in the GDR$\rightarrow $DGDR transition. This
negative result ensures that calculations, in which only transitions between
collective components of the GDR and DGDR are taken into account and which
are much easier to carry out, require no further corrections.

The main mechanism for the DGDR excitation in RHIC within a semiclassical
approach is a two-step process g.s.$\rightarrow $GDR$\rightarrow $DGDR \cite
{Ber88}. Corrections to the second-order perturbation theory arising from
coupled-channels were studied in Ref. \cite{Ber96a}. Although for grazing
impact parameters the coupled-channels calculation deviates on the 20\%
level from second-order perturbation theory, it makes only a small change to
the total cross sections. That is, for not too small impact parameters the
second-order perturbation theory works quite well. Indeed, this has been
observed recently in the analysis of the experiment on DGDR excitation in
lead projectiles impinging on different $Z$-targets \cite{Bor96}.

Thus, we can use for the excitation probability of the DGDR

\begin{eqnarray}
P_{DGDR}(E_f,\;b) &=&\frac 14\sum_{M_f}\Bigg| \sum_{i,M_i}a_{0(0)\rightarrow
1_i^{-}(M_i)}^{E1(\mu )}(E_i,\;b)  \nonumber \\
\times &&a_{1_i^{-}(M_i)\rightarrow [1^{-}\times 1^{-}]_f(M_f)}^{E1(\mu
^{\prime })}(E_f-E_i,\;b)\Bigg| ^2  \nonumber \\
&&  \label{cs}
\end{eqnarray}
where the index $i$ labels intermediate states belonging to the GDR, and $%
a_{J_1(M_1)\rightarrow J_2(M_2)}^{E1(\mu )}$ is the first-order E1
excitation amplitude for the transition $J_1(M_1)\longrightarrow J_2(M_2)$
in a collision with impact parameter $b$. For each state, $J$ and $M$ denote
the total angular momentum and the magnetic projection, respectively.

The amplitude $a_{J_1(M_1)\rightarrow J_2(M_2)}^{E1(\mu )}$ is given by

\begin{eqnarray*}
a_{J_1(M_1)\rightarrow J_2(M_2)}^{E1(\mu )}(E,\;b) &=&(J_1M_11\mu |J_2M_2) \\
\times  &<&J_2||E1||J_1>f_{E1(\mu )}(E,\;b)\;.
\end{eqnarray*}
It is a product of the reduced matrix element $<J_2||E1||J_1>$ for the $E1$%
-transition between the states $J_1(M_1)$ and $J_2(M_2)$ which carries
nuclear structure information and the reaction function $f_{E1(\mu )}(E,\;b)$%
. The latter depends on the excitation energy, charge of the target, beam
energy, and is calculated according to Ref. \cite{WA79}. Except for the
dependence on the excitation energy, it does not carry any nuclear structure
information. The cross section for the DGDR is obtained from Eq. (\ref{cs})
by integration over impact parameters, starting from a minimal value $b_{min}
$ to infinity. This minimal value is chosen according to Ref. \cite{Ber96a}.

In microscopic approaches the strength of the GDR is split among several
one-phonon $1_\alpha ^{-}$ states (due to the Landau damping). The wave
function $|1_\alpha ^{-}>$ couples to complex configurations $|1_\beta ^{-}>$
yielding the GDR width. We use the index $\alpha $ for simple configurations
and the index $\beta $ for complex ones, respectively. Thus, the wave
function of the $i^{th}$ $1^{-}$ state in the GDR energy region has the form:

\begin{equation}
|1_i^{-}>=\sum_\alpha S_i^{\mbox{\tiny GDR}}(\alpha )|1_\alpha
^{-}>+\sum_\beta C_i^{\mbox{\tiny GDR}}(\beta )|1_\beta ^{-}>  \label{gdr}
\end{equation}
where coefficients $S_i^{\mbox{\tiny GDR}}(\alpha )$ and $C_i^{%
\mbox{\tiny
GDR}}(\beta )$ can be obtained by diagonalizing the nuclear model
Hamiltonian on the set of wave functions (\ref{gdr}).

The total $E1$-strength of the GDR excitation from the ground state, $B_{%
\mbox{\tiny GDR}}(E1)=\sum_i|<1_i^{-}||E1||0_{g.s.}^{+}>|^2,$ remains
practically the same as in the one-phonon RPA calculation because the direct
excitation of complex configurations from the ground state is a few order of
magnitude weaker as compared to excitation of one-phonon states. However
these complex configurations play a fundamental role for the width of the
GDR.

The wave function of the $2^{+}$ component of the DGDR states can be written
in the similar fashion:

\begin{eqnarray}
|2_f^{+}>= &&\sum_{\tilde{\alpha}=\{\alpha _1\times \alpha _2\}}S_f^{%
\mbox{\tiny
DGDR}}(\tilde{\alpha})|[1_{\alpha _1}^{-}\times 1_{\alpha _2}^{-}]_{2^{+}}> 
\nonumber \\
+ &&\sum_{\alpha ^{\prime \prime }}{\tilde{S}}_f^{\mbox{\tiny DGDR}}(\alpha
^{\prime \prime })|2_{\alpha ^{\prime \prime }}^{+}>+\sum_{\beta ^{\prime
}}C_f^{\mbox{\tiny DGDR}}(\beta ^{\prime })|2_{\beta ^{\prime }}^{+}> 
\nonumber  \label{dgdr} \\
&&  \label{dgdr}
\end{eqnarray}
In this equation we separated in the first term the doorway $[1^{-}\times
1^{-}]$ DGDR configurations from other two-phonon configurations (second
term) and complex configurations (the last term). The same equation as (\ref
{dgdr}) is valid for the $0^{+}$ DGDR states (see Ref. \cite{Ber96} for a
discussion on the role of the $0^{+},\;1^{+},\;$and $2^{+}$ DGDR states in
the excitation process).

The total $E1$-transition strength between the GDR and DGDR, $%
\sum_f\sum_i|<2^{+}(0^{+})_f||E1||1_i^{-}>|^2,$ is much larger as compared
to that for the GDR excitation, $\sum_i|<1_i^{-}||E1||0_{g.s.}^{+}>|^2,$
from the ground state. This is because the former includes transitions not
only between doorway GDR and DGDR states but also between complex
configurations as well. The enhancement factor should be the ratio between
the density of doorway and complex configuration in the GDR energy region.
But in the two-step excitation process the sum over intermediate GDR states
in Eq.~(\ref{cs}) reduces the total transition strength for $g.s.\rightarrow
GDR\rightarrow DGDR$ to $\sim 2\cdot |B_{\mbox{\tiny GDR}}(E1)|^2$ (the
factor 2 appears due to the bosonic character of the two phonons which also
holds if Landau damping is taken into account). To prove this we substitute
the wave functions of the GDR and DGDR states given by Eqs.~(\ref{gdr},\ref
{dgdr}) in expression (\ref{cs}). We obtain two terms. The first one
corresponds to transitions between doorway GDR and DGDR states (after the
GDR is excited from the ground state through its doorway component):

\begin{eqnarray}
A_{\mu \mu ^{\prime }} &=&\sum_i\sum_{\alpha \alpha ^{\prime }\tilde{\alpha}%
}S_i^{\mbox{\tiny
GDR}}(\alpha )f_{E1(\mu )}(E_i,\;b)<1_\alpha ^{-}||E1||0_{g.s.}^{+}> 
\nonumber \\
\times  &&S_i^{\mbox{\tiny GDR}}(\alpha ^{\prime })S_f^{\mbox{\tiny DGDR}}({%
\tilde{\alpha}})f_{E1(\mu ^{\prime })}(E_f-E_i,\;b)  \nonumber \\
\times  &<&[1_{\alpha _1}^{-}\times 1_{\alpha _2}^{-}]_f||E1||1_{\alpha
^{\prime }}^{-}>\delta _{\alpha _2,\alpha ^{\prime }}  \label{sim}
\end{eqnarray}
and the second one accounts transitions between complex configurations in
the wave functions of Eqs.~(\ref{gdr},\ref{dgdr}):

\begin{eqnarray}
B_{\mu \mu ^{\prime }} &=&\sum_i\sum_{\alpha \alpha ^{\prime }\beta \beta
^{\prime }}S_i^{\mbox{\tiny GDR}}(\alpha )f_{E1(\mu )}(E_i,\;b)<1_\alpha
^{-}||E1||0_{g.s.}^{+}>  \nonumber \\
&&\times C_i^{\mbox{\tiny GDR}}(\beta )C_f^{\mbox{\tiny DGDR}}(\beta
^{\prime })f_{E1(\mu ^{\prime })}(E_f-E_i,\;b)  \nonumber \\
\times &<&[1_{\alpha ^{\prime }}^{-}\times 1_\beta ^{-}]_f||E1||1_\beta
^{-}>\delta _{\beta ^{\prime },[\alpha ^{\prime }\times \beta ]}~.
\label{com}
\end{eqnarray}

The second reduced matrix element in the above equations is proportional to
the reduced matrix element between the ground state and the doorway
one-phonon configuration \cite{Ber96}.

For a given impact parameter $b$, the function $f_{E1(\mu )}(E,\;b)$ can be
approximated by a constant value $f_{E1(\mu )}^0$ \cite{Ber88} for the
relevant values of the excitation energies. Then the energy dependence can
be taken out of summations and orthogonality relations between different
components of the GDR wave functions can be applied \cite{Pon96a}. The
orthogonality relations between the wave functions imply that $\sum_iS_i^{%
\mbox{\tiny
GDR}}(\alpha )C_i^{\mbox{\tiny
GDR}}(\beta )\equiv 0.$ This means that the term $B_{\mu \mu ^{\prime }}$
vanishes. The term $A_{\mu \mu ^{\prime }}$ summed over projections and all
final states yields a transition probability to the DGDR, $P_{DGDR}(E_f,\;b),
$ which is proportional to $2\cdot |B_{\mbox{\tiny GDR}}(E1)|^2$ in second
order perturbation theory. This argument was the reason for neglecting the
term $B_{\mu \mu ^{\prime }}$ in previous calculations of DGDR excitation in
Refs.~\cite{Pon97,Pon96a,Pon96b} where the coupling of doorway GDR and DGDR
states to complex configurations was taken into account.

In Fig.~\ref{fig1} we plot the value of $\chi _{E1}(E)=2\pi \int
db\;b\sum_\mu |f_{E1(\mu )}(E,\;b)|^2$ as a function of energy calculated
for the $^{208}$Pb (640$\cdot $A~MeV)~+~$^{208}$Pb reaction. This value
corresponds to $\sigma _{\mbox{\tiny GDR}}$ if $B_{\mbox{\tiny GDR}}(E1)=1$.
The square in this figure indicates the location of the DGR in $^{208}$Pb.
This figure demonstrates that the function $\chi _{E1}(E)$ changes by 60\%
in the GDR energy region. The role of this energy dependence for other
effects has been considered in Refs.~\cite{Ber96a,Ber96,Lan97}. Taking into
account that one-phonon $1_\alpha ^{-}$ configurations are fragmented over a
few MeV \cite{Pon96b}, when a sufficiently large two-phonon basis is
included in the wave function given by Eq.~(\ref{gdr}), the role of the $%
B_{\mu \mu ^{\prime }}$ term in the excitation of the DGDR should be studied
in more detail.

To accomplish this task we have performed firstly a simplified calculation
in which we used the {\it boson type }Hamiltonian:

\begin{eqnarray}
H &=&\sum_\alpha \omega _\alpha Q_\alpha ^{\dagger }Q_\alpha +\sum_\beta 
\widetilde{\omega }_\beta \widetilde{Q}_\beta ^{\dagger }\widetilde{Q}_\beta
\nonumber \\
&&+\sum_{\alpha ,\beta }U_\beta ^\alpha (Q_\alpha ^{\dagger }\widetilde{Q}%
_\beta +h.c.)  \label{h}
\end{eqnarray}
where $Q_\alpha ^{\dagger }$ is the phonon creation operator and $\omega
_\alpha $ is the energy of this one-phonon configuration; $\widetilde{Q}%
_\beta ^{\dagger }$ is the operator for creation of a complex configuration
with energy $\widetilde{\omega }_\beta $ and $U_\beta ^\alpha $ is the
matrix element for the interaction between these configurations. We have
assumed that the energy difference between two neighboring one-phonon
configurations is constant and equals to $\Delta \omega $. An equidistant
spacing with the energy $\Delta \widetilde{\omega }$ was assumed for the
complex configurations. We also have used a constant value $U$ for the
matrix elements of the interaction. The $B_{\mbox{\tiny GDR}}(E1)$ value was
distributed symmetrically over doorway one-phonon configurations. Thus, the
free parameters of this model are: $\Delta \omega $, $\Delta \widetilde{%
\omega }$, $U$, the number of one-phonon and complex configurations, and the
distribution of the $B_{\mbox{\tiny GDR}}(E1)$ value among the doorway
states. The only condition we want to be satisfied is that the energy
spectrum for the GDR photoexcitation is the same as the one known from the
experiment.

After all parameters are fixed we diagonalize the model Hamiltonian of Eq.~(%
\ref{h}) on the set of wave functions of Eq. (\ref{gdr}) for the GDR and on
the set of Eq. (\ref{dgdr}) for the DGDR. The diagonalization procedure
yields the information on eigen energies of the $1_i^{-}$ GDR states and on
the coefficients $S_i^{\mbox{\tiny GDR}}(\alpha )$ and $C_i^{\mbox{\tiny GDR}%
}(\beta ),$ respectively. One also obtains information on eigen energies of
the $2_f^{+}$ or $0_f^{+}$ DGDR states and the coefficients $S_f^{%
\mbox{\tiny
DGDR}}(\tilde{\alpha})$ and $C_f^{\mbox{\tiny DGDR}}(\beta ^{\prime }),$
respectively. With this information we are able to study the role of the $%
B_{\mu \mu ^{\prime }}$ term in the excitation of the DGDR in RHIC.

The big number of free parameters allows an infinite number of suitable
choices. In fact, not all of the parameters are really independent. For
example, the increase in the number of simple or complex configurations goes
together with the decreasing of the value of $U.$ This is necessary for a
correct description of the GDR photoabsorption cross section. This makes it
possible to investigate the role of the $B_{\mu \mu ^{\prime }}$ term in
different conditions of weak and strong Landau damping and for different
density of complex configurations. In our calculations we vary the number of
collective doorway states from one to seven and the number of complex
configurations from 50 to 500. The value of $U$ then changes from about 100
to 500~keV. The results of one of these calculations for the excitation of
the $2^{+}$ component of the DGDR in $^{208}$Pb (640$\cdot $A~MeV)~+~$^{208}$%
Pb collisions are presented in Fig.~\ref{fig2}. For a better visual
appearance the results are averaged with a smearing parameter equal to
1~MeV. The dashed curve shows the results of a calculation in which $\sigma
_{\mbox{\tiny DGDR}}^A(E)\equiv \sigma _{\mbox{\tiny DGDR}}(E)\sim \int
db\;b|A_{\mu \mu ^{\prime }}|^2$ and the results of another one in which $%
\sigma _{\mbox{\tiny DGDR}}^{A+B}(E)\equiv \sigma _{\mbox{\tiny
DGDR}}(E)\sim \int db\;b\left| A_{\mu \mu ^{\prime }}+B_{\mu \mu ^{\prime
}}\right| ^2$ are represented by a solid curve.

Our calculation within this simple model indicates that the role of the $%
B_{\mu \mu ^{\prime }}$ term in the second order perturbation theory is
negligibly small, although the total $B(E1)$ strength for transitions
between complex GDR and DGDR configurations, considered separately, is more
than two orders of magnitude larger than the ones between doorway GDR and
DGDR configurations. The value $\Delta \sigma =(\sigma _{\mbox{\tiny DGDR}%
}^{A+B}-\sigma _{\mbox{\tiny DGDR}}^A)/\sigma _{\mbox{\tiny DGDR}}^A$, where 
$\sigma _{\mbox{\tiny DGDR}}^{A(A+B)}=\int \sigma _{\mbox{\tiny
DGDR}}^{A(A+B)}(E)dE$, changes in these calculations from 1\% to 2.5\%. The
results practically do not depend on the number of complex configurations
accounted for. The maximum value of $\Delta \sigma $ is achieved in a
calculation with a single doorway GDR state (no Landau damping). This is
because the value of $U$ is the larger in this case and the fragmentation of
the doorway state is stronger. Thus, in such a situation, the energy
dependence of the reaction amplitude modifies appreciably the orthogonality
relations. But in general the effect is marginal.

We also performed a calculation with more realistic wave functions for the
GDR and DGDR states taken from our previous studies \cite{Pon97} which were
based on microscopic QPM (Quasiparticle Phonon Model) \cite{Sol92}. These
wave functions include 6 and 21 doorway states for the GDR and DGDR,
respectively. The complex configurations are two-phonon states for the GDR
and three-phonon states for the DGDR. The energies of the doorway states and
complex configurations were obtained from RPA equations and the matrix
elements of the interaction, $U_\beta ^\alpha ,$ were calculated on a
microscopic footing without free parameters by making use of the QPM
Hamiltonian and internal fermion structure of phonons. The value $\Delta
\sigma $ equals in this realistic calculation to 0.5\%. This result is not
surprising because realistic calculations with only two-phonon complex
configurations, and a limited number of them, somewhat underestimate the GDR
width which is crucial for the modification of the orthogonality relations.

In conclusion, we investigated the role of transitions between complex GDR
and DGDR configurations within second-order perturbation theory for the DGDR
excitation in RHIC. We have proved that these transitions play a marginal
role in the process under consideration and it is sufficient to take into
account only transitions between the ground state and doorway GDR and DGDR
configurations.

V.Yu. P. thanks for the hospitality at the Instituto de F\'{\i}sica of the
Universidade Federal do Rio de Janeiro, where this research has been
performed, the CNPq for the financial support and Prof. P.F. Bortignon for
fruitful discussions. This work was partially supported by the RFBR (grant
96-15-96729), by the Heisenberg-Landau program, by the FUJB/UFRJ and by the
MCT/FINEP/CNPq(PRONEX) (contract 41.96.0886.00)

\begin{figure}[tbp]
\caption{The energy dependence of the $^{208}$Pb (640$\cdot $A~MeV)~+~$%
^{208} $Pb reaction function calculated within first order perturbation
theory. The square indicates the location of the GDR in $^{208}$Pb.}
\label{fig1}
\end{figure}

\begin{figure}[tbp]
\caption{The cross section for the excitation of the $2^{+}$ component of
the DGDR in the reaction $^{208}$Pb (640$\cdot $A~MeV)~+~$^{208}$Pb,
calculated within second order perturbation theory. The dashed curve shows
the contribution of the $E1$-transition between doorway GDR and DGDR
configurations only. The solid curve is a sum of the above result and the
contribution of the $E1$-transitions between complex GDR and DGDR
configurations. See text for details. }
\label{fig2}
\end{figure}

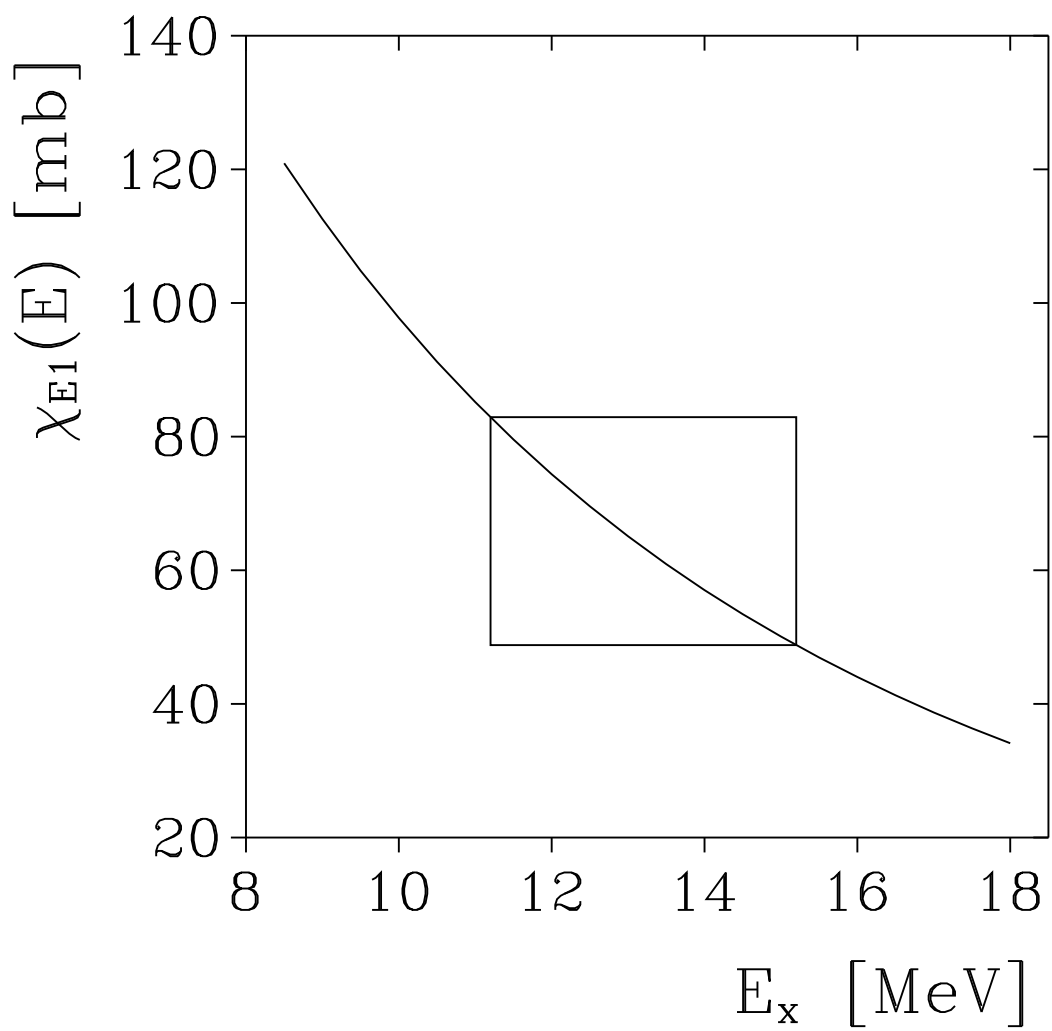
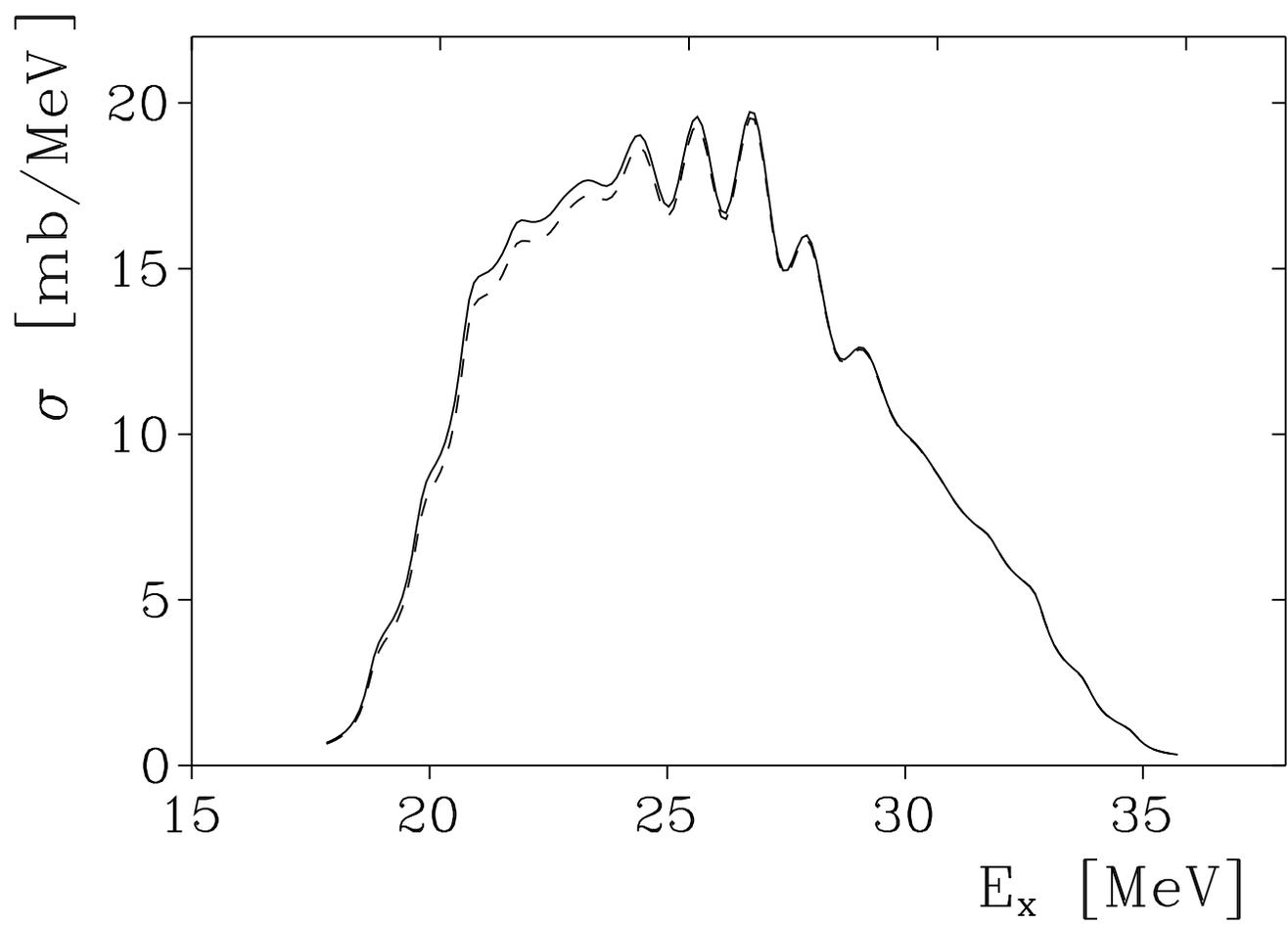
\end{document}